\documentclass[twoside]{elsart}
\usepackage{epsfig}
\usepackage{array}
\usepackage{amssymb,amsbsy,amsmath}
\usepackage{times}
\newcommand{\figref}[1]{Fig.~\protect\ref{#1}}
\newcommand{\secref}[1]{Sec.~\protect\ref{#1}}
\renewcommand{\eqref}[1]{Eq.~(\protect\ref{#1})}
\newcommand{\xref}[1]{\protect\ref{#1}}
\newcommand{\fmref}[1]{(\protect\ref{#1})}

\def\V0{\stackrel{\circ}{V}}
\def\v0{\stackrel{\circ}{v}}
\def\half{{\frac{1}{2}}}
\def\threehalf{\frac{3}{2}\;}

\newcommand{\dint}{\mbox{d}}
\renewcommand{\Re}[1]{\mbox{Re}\left(#1\right)}

\newcommand{\tr}{\mbox{tr}}
\newcommand{\op}[1]{%
    \fontdimen12\textfont3=2pt\fontdimen12\scriptfont3=1.4pt%
    \!\null\mathop{\vphantom{#1}\smash{#1}}\limits_{\sim}\null\!}

\def\bra#1{\langle \, {#1} \, | \,}
\def\ket#1{\, | \, {#1} \, \rangle}

\def\ketsm{\ket{s\,m}}
\def\brasm{\bra{s\,m}}

\def\ketsm1{\ket{s_1\,m_1}}
\def\brasm1{\bra{s_1\,m_1}}
\def\ketsm2{\ket{s_2\,m_2}}
\def\brasm2{\bra{s_2\,m_2}}
\newcommand{\pp}[2]{\frac{\partial \, {#1}}{\partial \, {#2}}\;}

\newcommand{\vek}[1]{{\!\vec{\,#1}}}
\begin{document}
%
\typeout{   --- >>>   trimer paper   <<<   ---   }
\typeout{   --- >>>   trimer paper   <<<   ---   }
\typeout{   --- >>>   trimer paper   <<<   ---   }
%
%
\journal{Physica A}
\begin{frontmatter}
\title{Transition from quantum to classical Heisenberg trimers: Thermodynamics and
time correlation functions}
 
\author{D. Mentrup, H.-J. Schmidt, J. Schnack\thanksref{JS}}
\address{Universit\"at Osnabr\"uck, Fachbereich Physik \\ 
         Barbarastr. 7, 49069 Osnabr\"uck, Germany}
\author{Marshall Luban}
\address{Ames Laboratory \& Department of Physics and Astronomy,
Iowa State University\\ Ames, Iowa 50011, USA}
\thanks[JS]{corresponding author: jschnack\char'100uos.de,\\ http://www.physik.uni-osnabrueck.de/makrosysteme/}

\begin{abstract}
\noindent
We focus on the transition from quantum to classical behavior in
thermodynamic functions and time correlation functions of a
system consisting of three identical quantum spins $s$ that
interact via isotropic Heisenberg exchange. The partition
function and the zero-field magnetic susceptibility are readily
shown to adopt their classical forms with increasing $s$. The
behavior of the spin autocorrelation function (ACF) is more
subtle. Unlike the classical Heisenberg trimer where the ACF
approaches a unique non-zero limit for long times, for the
quantum trimer the ACF is periodic in time. We present exact
values of the time average over one period of the quantum trimer
for $s\le 7$ and for infinite temperature. These averages differ
from the long-time limit, $(9/40)\ln3 + (7/30)$, of the
corresponding classical trimer by terms of order
$1/s^2$. However, upon applying the Levin $u$-sequence
acceleration method to our quantum results we can reproduce the
classical value to six significant figures.

\vspace{1ex}

\vspace{1ex}

\end{abstract}
\end{frontmatter}
\raggedbottom
\section{Introduction and summary}

In recent years considerable attention has been devoted to the magnetic
properties of synthesized organic complexes (``molecular magnets")
containing small numbers of paramagnetic ions
\cite{BeG90,GCR94,Gat94,FST96,PDK97}. With the ability to
control the placement of magnetic moments of diverse species
within stable molecular structures, it is now possible to test
basic questions concerning their magnetic properties and to
explore the design of novel systems that offer the prospect of
useful applications \cite{Her93,BuP95}. Intermolecular magnetic
interactions are typically extremely weak compared to
intramolecular interactions, so a bulk sample can be considered
as independent individual molecular magnets.

The present study is motivated by the successful synthesis of
two trimers, one \cite{MMB98} consisting of V$^{4+}$ ions ($s=1/2$)
and the second \cite{CCF95} consisting of Fe$^{3+}$ ions
($s=5/2$). With the anticipation of the successful synthesis of
yet other trimers, in this article we consider the thermodynamic
functions and the time correlation functions for three identical
quantum spins $s$ which interact via isotropic Heisenberg
exchange in the absence of an external magnetic field. We are
especially interested in comparing the behavior of these
quantities for diverse $s$, and in particular in the transition
to classical behavior which occurs for large $s$. This
transition is readily analyzed for the thermodynamic
functions. In particular, we show that the partition function
and the zero-field magnetic susceptibility adopt the correct
classical forms \cite{CLA99} with increasing $s$. Our analysis
also provides results for the deviation from classical behavior
depending on the size of $s$ and the temperature of the system.

Knowledge of the exact two-spin time correlation functions is of
great importance since this enables one to predict the outcome
of measurements such as the proton-spin lattice relaxation rate
\cite{Mor56} by nuclear magnetic resonance (NMR) techniques
and inelastic magnetic neutron scattering \cite{BaL89}.  In an
earlier publication \cite{MSL99} we examined in detail the
properties of the thermal equilibrium autocorrelation function,
to be denoted by $A_s(t,T)$, for a quantum Heisenberg dimer
composed of two identical spins $s$ which interact with
isotropic exchange.  Some features of $A_s(t,T)$ that occur for
a dimer are readily shown to occur for the trimer, too, and we
will not deal with those issues. However, there are two features
of $A_s(t,T)$ for the quantum trimer which deserve careful
attention and these are considered in detail.

First, for the quantum Heisenberg trimers the quantity
$A_s(t,T)$ is a periodic function of the time with a recurrence
time $\tau$. For the corresponding classical Heisenberg trimer,
whose ACF is denoted by $A_c(t,T)$, there exists a unique,
non-zero long-time limit $A_c(\infty,T)$ \cite{LBC99}.  To make
a meaningful comparison between the asymptotic, long-time
behavior of the classical Heisenberg trimer with a quantum
trimer, we suggest comparing $A_c(\infty,T)$ with the time
average of $A_s(t,T)$ over the corresponding recurrence time
$\tau$, to be denoted by $\overline{A}_s(T)$; this value is
equivalent to the coefficient of $\delta(\omega)$ in the
expression for the Fourier time transform of $A_s(t,T)$. In
particular we will explore the large $s$ behavior of
$\overline{A}_s(T)$. Second, for the classical Heisenberg
trimer, and for the special case of infinite temperature it has
been shown that $A_c(\infty,\infty)=(9/40) \ln 3 + 7/30$ \cite{LBC99}.
This curious exact value differs from the result, $1/N$, that
follows from a phenomenological diffusive spin dynamics (DSD)
\cite{LBC99} based on linear equations of motion for a ring of
$N$, here $N=3$, interacting classical Heisenberg spins.  By
contrast, for both the quantum and classical dimers it has been
found that $\overline{A}_s(\infty)=A_c(\infty,\infty)=\half$,
independent of $s$ and in agreement with the DSD result. It is
thus of interest to track the emergence of the exact classical
result upon considering the sequence of quantum trimers for
increasing $s$.

We have derived the exact values of $\overline{A}_s(\infty)$ for
the quantum trimers for the choices $s=1/2,1,3/2,\dots,7$. It is
necessary to consider the results for half-integer $s$
separately from those derived for integer $s$ since they exhibit
different behaviour patterns. Each subsequence appears to
converge very slowly to the above value of $A_c(\infty,\infty)$
for the classical trimer. For spins $s$ the deviation of
$\overline{A}_s(\infty)$ from the classical result is of order
$1/s^2$. However, when we apply the Levin $u$-sequence
acceleration method \cite{Lev73,Lub77} to the subsequence for
half-integer $s$ we arrive at an estimate for the
large-$s$-limit which agrees to 6 significant figures with
$A_c(\infty,\infty)$. For integer $s$ the Levin $u$-estimate
agrees with $A_c(\infty,\infty)$ to 5 significant figures.

The layout of this paper is as follows. In \secref{sec-2} we
calculate the partition function and the zero-field
susceptibility of the quantum trimer for general $s$. We show
that both quantities approach the corresponding results for the
classical trimer. In \secref{sec-3} we derive a general formula
for the time average $\overline{A}_s(\infty)$. Numerical
evaluation of the formula becomes a very lengthy process with
increasing $s$, however, we have been able to perform these
calculations for $s\le 7$. The Levin $u$-estimates for
$\overline{A}_s(\infty)$ are also provided in \secref{sec-3}.

\section{Thermodynamic functions}
\label{sec-2}

The quantum trimer is specified by the Hamilton operator
\begin{eqnarray}
\label{E-2-1}
\op{H}_0
&=&
\frac{J}{\hbar^2}\;
\left(
\op{\vek{s}}_1 \cdot \op{\vek{s}}_2
+
\op{\vek{s}}_2 \cdot \op{\vek{s}}_3
+
\op{\vek{s}}_3 \cdot \op{\vek{s}}_1
\right)
\\
&=&
\frac{J}{2 \hbar^2}\;
\left(
\op{\vek{S}}^2 - \op{\vek{s}}_1^2 - \op{\vek{s}}_2^2 - \op{\vek{s}}_3^2
\right)
\quad\ ;\quad
\op{\vek{S}}=\op{\vek{s}}_1 + \op{\vek{s}}_2 + \op{\vek{s}}_3
\ ,
\nonumber
\end{eqnarray}
where the spin operators satisfy the usual commutation relations
and where $J$ has units of energy. $J>0$ describes
antiferromagnetic and $J<0$ ferromagnetic coupling. Throughout
this article it is assumed that the spin quantum numbers of the
three sites are identical, $s_1=s_2=s_3=s$.  The eigenstates
$\ket{S,M,S_{23}}$ of the Hamilton operator can be chosen as
simultaneous eigenstates of the total spin $\op{\vek{S}}^2$, its
$z$-component $\op{{S}}_z$, and of
$\op{\vek{S}}_{23}^2=(\op{\vek{s}}_2+\op{\vek{s}}_3)^2$. The
quantum numbers $S,M,S_{23}$ characterize the eigenstates
completely.  The eigenvalues $E_S$ of the Hamilton operator
\begin{eqnarray}
\label{E-2-3}
E_S
&=&
\frac{J}{2}\;
\left(
S (S+1) - 3\,s (s+1)
\right)
\end{eqnarray}
do not depend on $S_{23}$ or $M$.
Thus the partition function in the canonical ensemble reads
\begin{eqnarray}
\label{E-2-4}
Z
&=&
\tr\left\{ e^{-\beta\op{H}_0}\right\} 
=
\sum_{S,M,S_{23}} \bra{S,M,S_{23}} e^{-\beta\op{H}_0} \ket{S,M,S_{23}}
\ .
\end{eqnarray}
The respective classical Hamilton function $H_{c}$ is defined as
\cite{MSL99}
\begin{eqnarray}
\label{E-2-5}
H_{c}
=
J_{c}\;
\left(
 \vek{e}_1 \cdot \vek{e}_2
+\vek{e}_2 \cdot \vek{e}_3
+\vek{e}_3 \cdot \vek{e}_1
\right)
\quad , \quad
J_{c} = J\,s(s+1)
\ ,
\end{eqnarray}
where $\vek{e}_1$, $\vek{e}_2$ and $\vek{e}_3$ are unit vectors
(c-numbers).  Then the classical partition function turns out to
be \cite{CLA99}
\begin{eqnarray}
\label{E-2-6}
Z_c
=
\int\, \dint E\, D_c(E) \exp(-\beta E)
\ ,
\end{eqnarray}
with the classical density of states
\begin{equation}
\label{E-2-7}
D_c(E)
=
\left\{ \begin{array}{l@{\quad:\quad}l} \vspace{2mm}
\frac{1}{2 J_c} \sqrt{\frac{2E}{J_c} +3}   & -\threehalf J_c \le E \le -J_c \\
\vspace{2mm}
\frac{1}{4 J_c} (3-\sqrt{\frac{2E}{J_c}+3})& -J_c < E \le 3 J_c\\
0 & \mbox{else} \end{array} \right.
\ ,
\end{equation}
which is normalized to unity.
In order to compare quantum and classical density of states, the
energy spectra of the quantum trimers for different $s$ have to
be mapped onto the same energy interval; we take $[-3 J_c/2, 3
J_c]$, i.e. all energies are divided by $s(s+1)$.

\begin{figure}[t]
\begin{center}
\epsfig{file=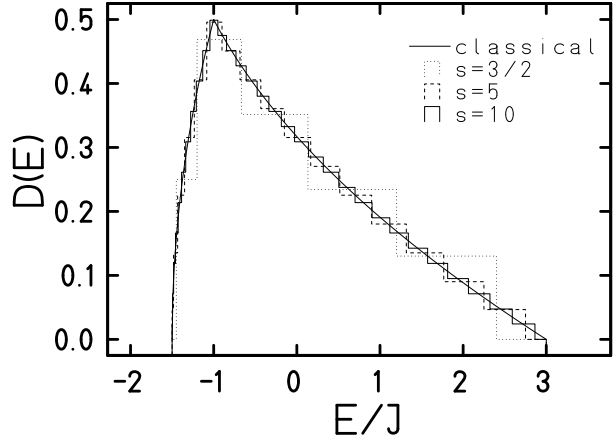,width=80mm}
\vspace*{1mm}
\caption[]{Normalized density of states for quantum Heisenberg
trimers (dashed lines) and their classical counterpart (solid line).}
\label{F-2-2}
\end{center} 
\end{figure} 

Figure \xref{F-2-2} demonstrates how the quantum density of
states approaches the classical limit with increasing spin
$s$. Considering the eigenvalues \fmref{E-2-3} of the Hamilton
operator \fmref{E-2-1} and their multiplicities, which originate
from the degeneracy in the magnetic quantum number $M$ and from
different possibilities to couple to a certain total spin $S$,
one can show analytically that the quantum density of states
converges against the classical one for infinite $s$
\cite{Men99}.

\begin{figure}[t]
\begin{center}
\epsfig{file=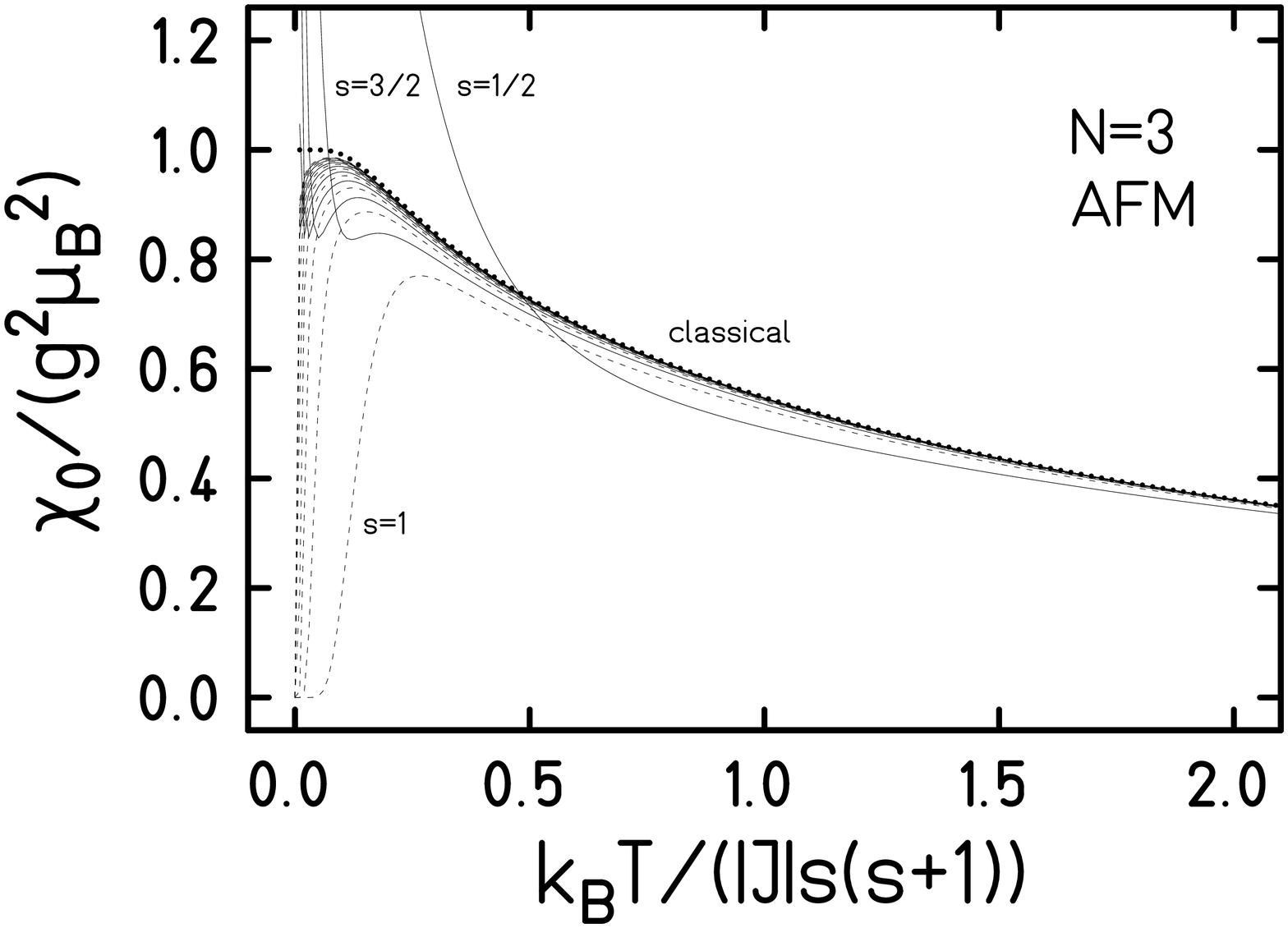,width=65mm}
$\quad$
\epsfig{file=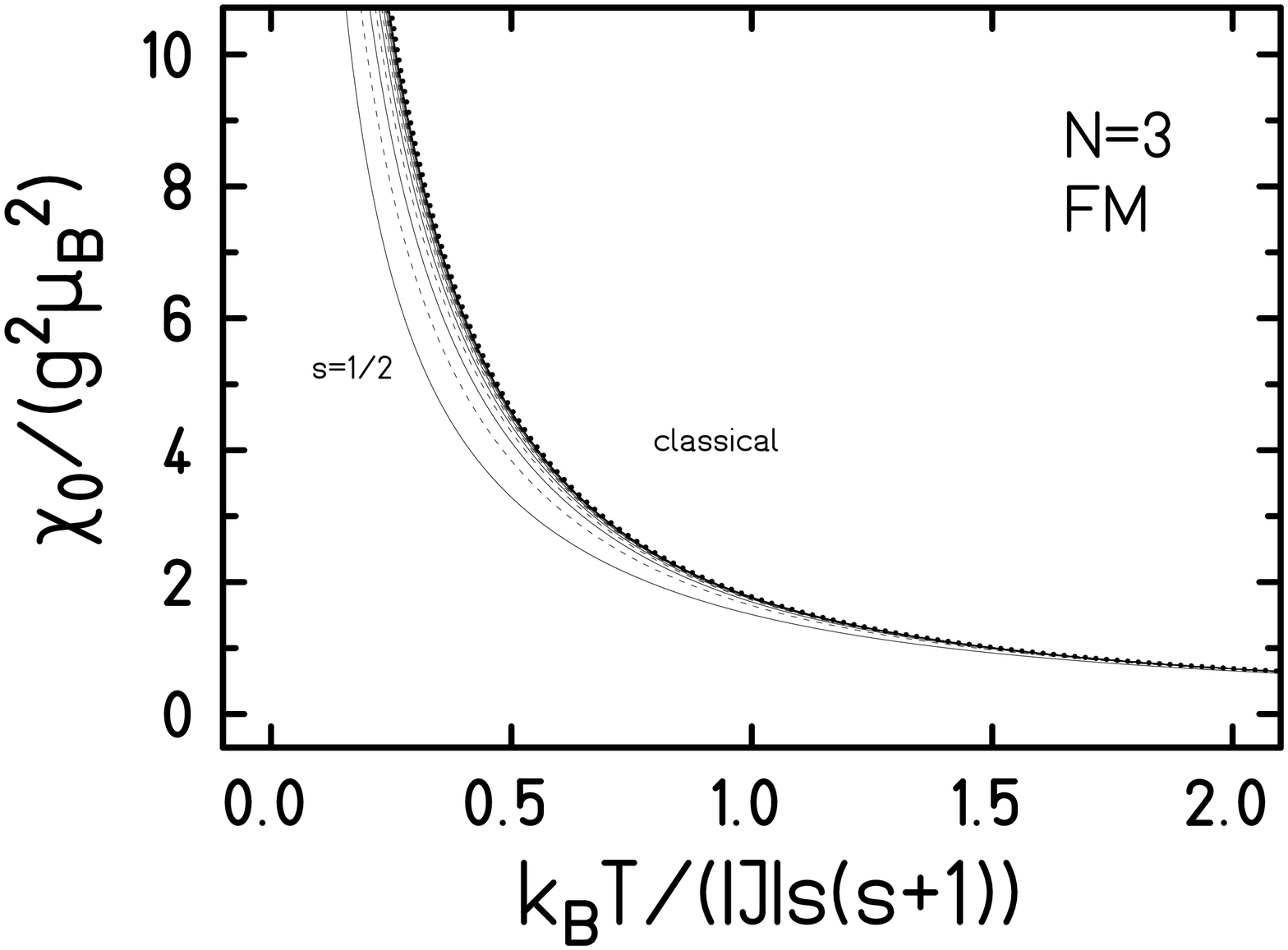,width=65mm}
\vspace*{1mm}
\caption[]{Zero-field susceptibility for $N=3$ and choices
$s=1/2,1,\dots,9$. The solid lines show the result for half
integer spins, the dashed lines for integer spin quantum
numbers. The classical result is given by the thick dotted line.
L.h.s.: antiferromagnetic coupling, r.h.s.: ferromagnetic coupling.}
\label{F-2-1}
\end{center} 
\end{figure}

But the coincidence of the classical and the quantum density of
states for high spin does not mean that other observables
coincide, too. An interesting example is given by the zero field
susceptibility. The susceptibility is defined as the derivative
of the magnetisation
\begin{eqnarray}
\label{E-2-8}
{\mathcal M}
&=&
\frac{1}{Z}\,
\tr\left\{-g \mu_B \op{S}_z \mbox{e}^{-\beta\op{H}}\right\}
\quad , \quad
\op{H}
=
\op{H}_0
+
g \mu_B B\op{{S}}_z
\end{eqnarray}
with respect to the magnetic field $B$
\begin{eqnarray}
\label{E-2-9}
\chi_0
&=&
\left(\pp{{\mathcal M}}{B}\right)_{B=0}
=
g^2 \mu_B^2 \beta
\left(
\frac{1}{Z}\,
\tr\left\{\left(\op{S}_z\right)^2
\mbox{e}^{-\beta\op{H}}\right\}
\right)_{B=0}
\ .
\end{eqnarray}
For ferromagnetic coupling, \figref{F-2-1} r.h.s., the graphs
for different $s$ nearly coincide with each other and with the
classical result. However, for antiferromagnetic coupling the
zero field susceptibility behaves differently for integer and
half-integer spin quantum numbers as shown on the l.h.s. of
\figref{F-2-1}. The reason is that for integer spin the ground
state, which is non-degenerate, has $S=0$ and thus the zero
field susceptibility approaches zero for small temperatures $T$,
whereas for half-integer spins the ground state, which is
fourfold degenerate, has $S=1/2$ which causes the susceptibility
to go to infinity for small $T$. However, if we consider a fixed
nonzero value of $T$, the susceptibilities tend to the classical
result with increasing $s$.  The classical result is
``indifferent" at $T=0$, it approaches $1$.

\section{Autocorrelation function}
\label{sec-3}

Another important observable is the two-spin time correlation
function because it serves as a major ingredient for several
quantities such as the spin lattice relaxation rate and the
neutron scattering cross section.

Considering that the Hamilton operator \fmref{E-2-1} is
isotropic, one obtains for the autocorrelation function
\begin{eqnarray}
\label{E-3-3}
A_s(t,T)
&=&
\frac{
\Re{
\sum_{S,M,S_{23}} \bra{S,M,S_{23}}\op{{s}}_{1z}(t) \cdot
\op{{s}}_{1z}(0)\, e^{-\beta\op{H}_0} \ket{S,M,S_{23}}}
}{
\sum_{S,M,S_{23}} \bra{S,M,S_{23}}\op{{s}}_{1z}(0) \cdot
\op{{s}}_{1z}(0)\, e^{-\beta\op{H}_0} \ket{S,M,S_{23}}
}
\ .
\end{eqnarray}

\begin{figure}[t]
\begin{center}
\epsfig{file=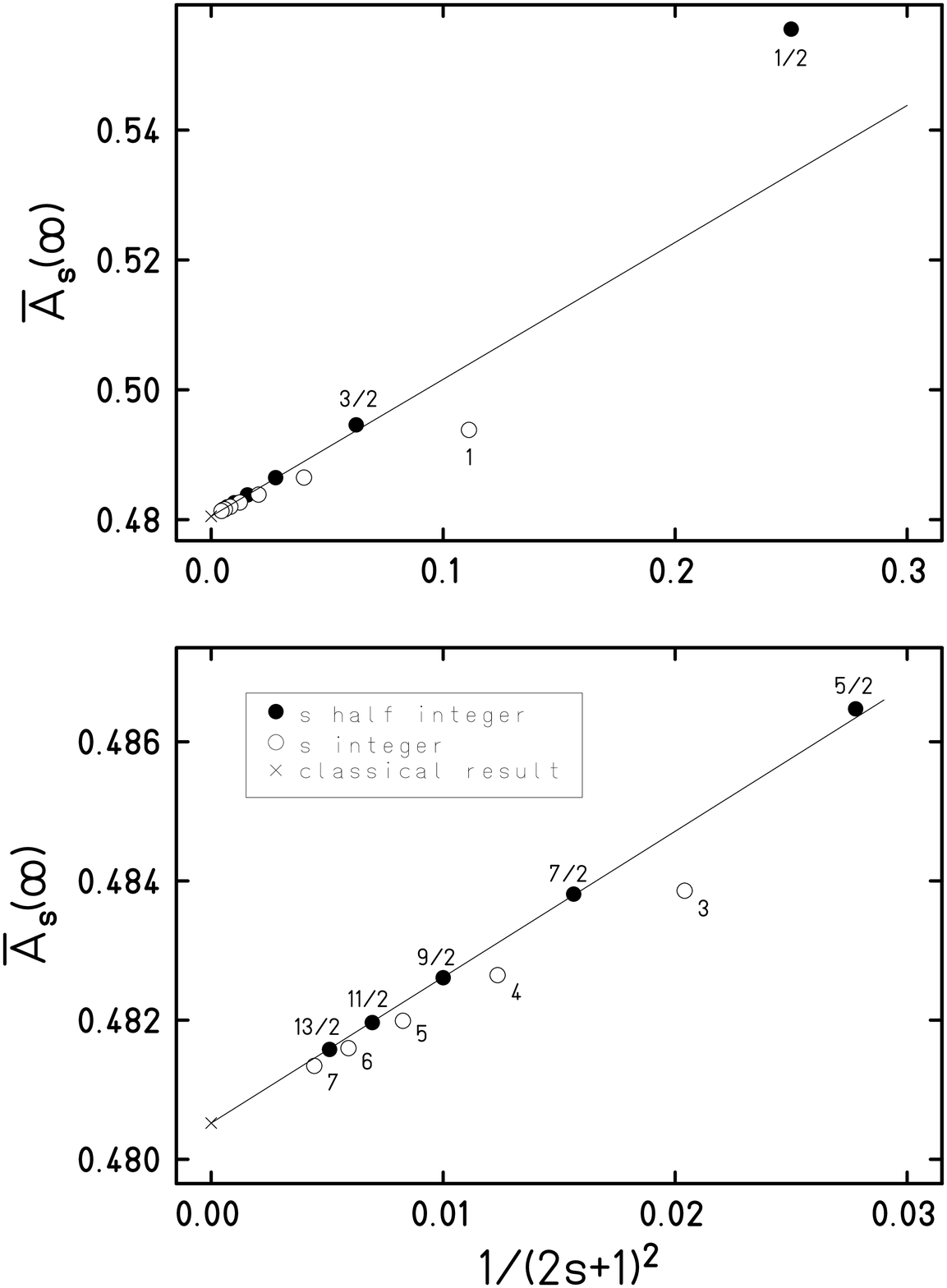,height=120mm}
\caption{Results for $\overline{A}_s(\infty)$ (open and full
circles) and $A_c(\infty,\infty)$ (cross). The line, which
connects the results for $s=11/2$ and $s=13/2$, is drawn to
guide the eye.}
\label{F-3-1}
\end{center} 
\end{figure} 

Since
$\bra{S,M,S_{23}}\op{{s}}_{1z}\ket{S^{\prime},M^{\prime},S_{23}^{\prime}}$
is zero if $|S-S^{\prime}|>1$, the contributing frequencies are
$\omega=J(S+1)/\hbar$. They are all multiples of a basic
frequency, which is $\omega_r=J/\hbar$ for integer spin quantum
numbers and $\omega_r=J/(2\hbar)$ for half integer spin quantum
numbers. Thus the autocorrelation function is periodic with a
recurrence time $2\pi/\omega_r$.

We have evaluated the expression $\overline{A}_s(\infty)$ 
\begin{eqnarray}
\label{E-3-4}
\overline{A}_s(\infty)
&=&
\lim_{T\rightarrow\infty}
\int_0^{\frac{2\pi}{\omega_r}} \dint t\;
A_s(t,T)
\end{eqnarray}
for half-integer values $1/2 \le s \le 13/2$ and for
integer values $1\le s \le 7$, and these are listed in table
\xref{T-3-1}. It appears to be impractical to extend these results to
larger spin values since the amount of computer time required
grows at an astonishing rate with increasing $s$. Fortunately
additional calculations are unwarranted. In \figref{F-3-1} we
display our results versus the independent variable $1/(2s+1)^2$
along with the solid line which has been chosen to pass through
the quantum results for s=11/2 and s=13/2.
The good agreement between the results for
the larger half-integer values of $s$ and the solid line is
consistent with the conclusion that the deviation between the
quantum results and the classical trimer decreases monotonically
to zero but very slowly, the deviation being of order
$1/(2s+1)^2$. In fact, if we approximate
$\overline{A}_s(\infty)$ by the form
\begin{eqnarray}
\label{E-3-1}
\overline{A}_s(\infty)
=
A + \frac{B}{(2s+1)^2}
\ ,
\end{eqnarray}
we may use our results for $s=11/2$ and $13/2$ to determine the
unknown parameters $A$ and $B$. In particular the result
$A=0.480511$ provides an estimate for
$lim_{s\rightarrow\infty}\overline{A}_s(\infty)$.  This result
is rather close to the exact classical result \cite{LBC99}
\begin{eqnarray}
\label{E-3-5}
A_c(\infty,\infty)
=
(9/40) \ln 3 + 7/30
=
0.4805210983
\ .
\end{eqnarray}
Adopting \eqref{E-3-1} for the case of integer values of $s$ and
using our results for $s=6$ and $s=7$ we find that $A=0.480575$.

We may obtain an improved estimate for
$lim_{s\rightarrow\infty}\overline{A}_s(\infty)$ by exploiting
the Levin $u$-sequence acceleration method \cite{Lev73,Lub77}
which is tailor-made for such slowly convergent, monotonic
sequences as the ones we face. If the sequence elements are
labelled $U_1, U_2, U_3,\dots$ (to be identified with
$\overline{A}_{1/2}(\infty),\overline{A}_{3/2}(\infty),\dots$)
and if we define the quantities $u_1=U_1, u_2=U_2-U_1,
u_3=U_3-U_2, \dots$ then the Levin $u$-estimate for
$lim_{n\rightarrow\infty}U_n$ based on employing the first $M$
values of $U_n$ is given by
\begin{eqnarray}
\label{E-3-2}
U[M]
=
\frac{\sum_{k=1}^M (-1)^{k-1}\binom{M}{k}k^{M-2}\frac{U_k}{u_k}}
     {\sum_{k=1}^M (-1)^{k-1}\binom{M}{k}k^{M-2}\frac{1}{u_k}}
\ .
\end{eqnarray}
We find that $U[7]=0.48052085\dots$. 
For the corresponding sequence of integer values of $s$ we find
that $U[7]=0.4805179\dots$. 

It is interesting to note that for half-integer values of $s$ the
Levin $u$-estimates are closer to the exact classical result than
those for integer $s$.

\begin{table}[hhhh]
\begin{center}
\begin{tabular}{|c|c||c|c|}
\hline
$s$ & $\overline{A}_s(\infty)$ &$s$ & $\overline{A}_s(\infty)$ \\
\hline\hline
1/2 & $\frac{5}{9}=0.55555556$ & 1 & $\frac{40}{81}=0.49382716$\\
3/2 & $\frac{779}{1575}=0.49460317$ & 2 & $\frac{5473}{11250}=0.48648889$\\
5/2 & $\frac{986093}{2027025}=0.48647303$ & 3 & $\frac{59747}{123480}=0.48385973$\\
7/2 & $\frac{21117673}{43648605}=0.48381095$ & 4 & $\frac{464441}{962280}=0.48264642$\\
9/2 & $\frac{302812778207}{627448696875}=0.48260962$ & 5 & $\frac{26536}{55055}=0.48199074$\\
11/2& $\frac{2796327017071}{5801928464475}=0.4819651$ & 6 & $\frac{33240299}{69020952}=0.48159723$\\
13/2& $\frac{19699872589701257}{40906818968140125}=0.48157919$ & 7 & $\frac{11459968711}{23808330000}=0.48134282$\\
\hline
\end{tabular}
\vspace*{5mm}
\end{center}
\caption{Time average, $\overline{A}_s(\infty)$, of the quantum
autocorrelation function in the high-temperature limit for
$s=1/2,1,\dots,7$.}\label{T-3-1}
\end{table}


\section*{Acknowledgments}
The authors thank T.~Bandos and K.~B\"arwinkel
for valuable discussions. The Ames Laboratory
is operated for the United States Department of Energy by Iowa
State University under Contract No. W-7405-Eng-82.


\end{document}